# Environmental dielectric screening effect on exciton transition energies in single-walled carbon nanotubes

Yutaka Ohno,[1, *] Shinya Iwasaki,[1] Yoichi Murakami,[2] Shigeru Kishimoto,[1] Shigeo Maruyama,[2] and Takashi Mizutani[1, †]

[1]*Department of Quantum Engineering, Nagoya University, Furo-cho, Chikusa-ku, Nagoya 464-8603, Japan*

[2]*Department of Mechanical Engineering, The University of Tokyo, 7-3-1 Hongo, Bunkyo-ku, Tokyo 133-8656, Japan*

(Dated: April 8, 2007)

## Abstract

Environmental dielectric screening effects on exciton transition energies in single-walled carbon nanotubes (SWNTs) have been studied quantitatively in the range of dielectric constants from 1.0 to 37 by immersing SWNTs bridged over trenches in various organic solvents by means of photoluminescence and the excitation spectroscopies. With increasing environmental dielectric constant ($\epsilon_{\text{env}}$), both $E_{11}$ and $E_{22}$ exhibited a redshift by several tens meV and a tendency to saturate at a $\epsilon_{\text{env}} \sim 5$ without an indication of significant ($n,m$) dependence. The redshifts can be explained by dielectric screening of the repulsive electron-electron interaction. The $\epsilon_{\text{env}}$ dependence of $E_{11}$ and $E_{22}$ can be expressed by a simple empirical equation with a power law in $\epsilon_{\text{env}}$, $E_{\text{ii}} = E_{\text{ii}}^{\infty} + A\epsilon_{\text{env}}^{-\alpha}$. We also immersed a sample in sodium-dodecyl-sulfate (SDS) solution to investigate the effects of wrapping SWNTs with surfactant. The resultant $E_{11}$ and $E_{22}$, which agree well with Weisman's data [Nano Lett. **3**, 1235 (2003)], are close to those of $\epsilon_{\text{env}}$ of 2. However, in addition to the shift due to dielectric screening, another shift was observed so that the $(2n+m)$-family patterns spread more widely, similar to that of the uniaxial-stress-induced shift.

## I. INTRODUCTION

Optical spectroscopy of single-walled carbon nanotubes (SWNTs) has received increasing attention, not only for assigning the chiral vector, (n,m), of the SWNTs[1-4] but also to study the physics of the one-dimensional excitons.[5-8] In paticular, photoluminescence (PL) and the excitation spectroscopies are the most widely used for characterizing SWNTs with resonant Raman scattering spectroscopy. Two-photon absorption spectroscopy[6] and Rayleigh scattering spectroscopy combined with TEM or SEM observation[7] are also powerful techniques to investigate excitonic states of SWNTs and the bundle effect.

Recently, the environmental effect is one of the topics most investigated for its effect on the optical properties of SWNTs.[9-15] It is known that the optical transition energies vary, depending on the kind of surrounding surfactant used to individualize SWNTs.[9,13] Lefebvre *et al.* have reported the optical transition energies of SWNTs bridging between two micropillars fabricated on a Si wafer show a blueshift as compared to the SDS-wrapped SWNTs.[12] We have also compared air-suspended SWNTs grown on a quartz substrate with a grating structure to SDS-wrapped SWNTs, and have shown that the energy differences between air-suspended and SDS-wrapped SWNTs depend on (n,m), especially on chiral angle and on type of SWNTs, type-I ($2n + m$ mod 3 = 1) or type-II ($2n + m$ mod 3 =2).[15] These energy variations due to environmental conditions have been thought to be produced by the difference in the dielectric constant of the surrounding materials. The energy of the many-body Coulomb interactions between carriers depends on the environmental dielectric constant ($\epsilon_{\rm env}$), because the electric force lines contributing to the Coulomb interactions pass through the matrix, as well as the inside of the SWNT.[16,17] Note that the environmental effect is caused not only by the dielectric screening of Coulomb interactions, but also by chemical[13,14] and mechanical conditions.[18] Finnie *et al.* have reported that gas adsorption affects the emission energy of SWNTs.[14]

Investigations into the environmental effect have not been comprehensive, despite its importance in understanding the optical properties of SWNTs. Quantitative and separate investigations are necessary to understand the contributions of various environmental factors. In this study, we have focused on the dielectric screening effect, which has been quantitatively investigated by immersing the SWNTs grown over trenches into various liquids with different $\epsilon_{\rm env}$, from 1.9 to 37, by means of photoluminescence and the excitation spectroscopies.

## II. EXPERIMENTAL

For the study on environmental effects, the SWNTs grown between two micro-pillars[12,19] or over a trench[15,20] are suitable because the environmental conditions can be controlled intentionally. The samples used in this work are SWNTs grown on a quartz substrate with periodic trenches, as shown in Fig. 1. Both the period and depth of the trenches are 2 $\mu$m. The trenches were formed by photolithography, Al-metal evaporation and lift-off, and subsequent reactive-ion etching of the quartz with the Al mask. The SWNTs were grown by alcohol catalytic chemical vapor deposition[21] at 700°C for 60 s, after spin coating of a water solution of Co acetate. By optimizing the growth condition, isolated bridging SWNTs can be formed over the trenches. The density of the SWNTs was $\sim$0.2 $\mu$m$^{-1}$ along the direction of the trench. Such low-density, isolated SWNTs were necessary to observe luminescence when the sample was immersed in liquids. On the other hand, in case of a sample with a relatively high-density of SWNTs as $\sim$2 $\mu$m$^{-1}$), PL intensity was degraded by immersing in liquids, even though strong PL was obtained in air. This is probably because the SWNTs form bundles with neighbors upon immersion in liquids, in the case of a sample with a high-density of SWNTs.

The sample was mounted in a vessel with a quartz window, and immersed in various organic solvents with $\epsilon_{\mathrm{env}}$ from 1.9 to 37 (see Table I). We also immersed a sample in 1-wt% D$_2$O solution of sodium-dodecyl-sulfate (SDS) to investigate the effect of wrapping with surfactant molecules. PL and the excitation spectra were measured using a home-made facility consisting of a tunable, continuous-wave Ti/Sapphire laser (710$\sim$850 nm), a monochromator with a focal length of 25 cm, and a liquid-N$_2$-cooled InGaAs photomultiplier tube (1000$\sim$1600 nm). The excitation wavelength was monitored by a laser wavelength meter. The diameter of the laser spot on the sample surface was $\sim$1 mm, so that an ensemble of many SWNTs was detected.

## III. RESULTS AND DISCUSSION

Figure 2 shows PL maps of SWNTs (a)$\sim$(g) in various organic solvents with different $\epsilon_{\mathrm{env}}$ and (h) in SDS solution. By immersing the sample in the organic solvents, the emission and excitation peaks, which respectively correspond to $E_{11}$ and $E_{22}$, showed redshifts and spectral broadening. In the PL map measured in SDS solution[22], the peak positions agree well with those of Weisman's empirical Kataura plot for SDS-wrapped SWNTs represented by crosses[3].

The $E_{22}$-$E_{11}$ plots in air ($\epsilon_{\text{env}}$ = 1.0), hexane (1.9), chloroform (4.8), and SDS solution are shown in Fig. 3 . Both of $E_{11}$ and $E_{22}$ showed redshifts with increasing $\epsilon_{\text{env}}$, without significant ($n,m$) dependence. The amounts of the redshifts are 33~49 meV for $E_{11}$ and 26~30 meV for $E_{22}$. The redshift with increasing $\epsilon_{\text{env}}$ is consistent with the theoretical work by Ando.[16] The optical transition energy in SWNTs is given by a summation of the bandgap and the exciton binding energies. When the $\epsilon_{\text{env}}$ increases, the Coulomb interaction is enhanced and the exciton binding energy decreases. This leads to a blueshift of the optical transition energy in the SWNT. It should be noted that the bandgap is renormalized by the electron-electron repulsion interaction, and consequently the change in the $\epsilon_{\text{env}}$ affects not only the exciton binding energy, but also the bandgap. According to theoretical work,[16,24] the repulsive electron-electron energy is larger in magnitude than the exciton binding energy. Therefore, if we consider that both energies show a similar dependence on $\epsilon_{\text{env}}$, the decrease in the electron-electron repulsion energy exceeds the decrease in the exciton binding energy when $\epsilon_{\text{env}}$ increases. This results in a redshift in optical transition energies. The redshifts observed in the present experiment are attributed to the decrease in the repulsion energy of the electron-electron interaction with an increase in the $\epsilon_{\text{env}}$. Note that even though the redshift with increasing $\epsilon_{\text{env}}$ is consistent with the theoretical studies, the amount of the redshift is much smaller than the calculations. This is probably because in the present experiments, the $\epsilon_{\text{env}}$ only outside of the SWNTs was varied, whereas the theoretical studies used a dielectric constant for the whole system.

$E_{11}$ and $E_{22}$ are plotted as a function of $\epsilon_{\text{env}}$ in Fig. 4. The energy shifts show a tendency to saturate at $\epsilon_{\text{env}} \sim 5$. The energy variations can be fitted to an empirical expression with a power law in $\epsilon_{env}$,

$$E_{\text{ii}} = E_{\text{ii}}^{\infty} + A\epsilon_{\text{env}}^{-\alpha} \tag{1}$$

where $E_{\text{ii}}^{\infty}$ corresponds to a transition energy when $\epsilon_{env}$ is infinity, $A$ corresponds to the maximum value of the energy change by $\epsilon_{env}$, and $\alpha$ is a fitting parameter. $A$ and $\alpha$ are 38 meV and 1.9 for $E_{11}$, and 29 meV and 1.6 for $E_{22}$, respectively, on average. Perebeinos *et al.* have reported power law scaling for excitonic binding energy by dielectric constant, where the scaling factor $\alpha$ is estimated to be $\sim$ 1.4 in the range of $\epsilon > 4$.[17] Although the power law expression of Eq. 1 is quite similar to the Perebeinos's scaling law for exciton binding energy, the present empirical expression is attributed to the scaling of electron-electron repulsion energy by $\epsilon_{\text{env}}$, rather than exciton (electron-hole) binding energy as described above. Quite recently, such a power-law-like-downshift in optical transition energy with $\epsilon_{\text{env}}$ has been obtained by theoretical calculations based on a tight-binding model.[23]

We have previously pointed out that the $E_{11}$ and $E_{22}$ varies depending on $(n,m)$, in particular on chiral angle and on the type of SWNTs (type-I or type-II), comparing those of SDS-wrapped SWNTs to those of air-suspended SWNTs. The same behavior occurred by immersing a sample in SDS solution, as shown in Fig. 2(h). Most $E_{11}$ and $E_{22}$ of SDS-wrapped SWNTs show a redshift, except for $E_{22}$ of near-zigzag type-II SWNTs, which show a blueshift as compared to those in air. This behavior is inconsistent with the results of the $\epsilon_{\text{env}}$ dependence as described above, in which $(n,m)$ dependence is not significant. Comparing the PL maps in SDS solution to those in organic solvents as shown in Fig. 3, the equivalent $\epsilon_{\text{env}}$ of SDS-wrapped SWNTs would be $\sim 2$. In addition to the shift due to the dielectric screening, the peak positions of SDS-wrapped SWNTs shift so as the $(2n+m)$-family patterns spread more widely. This suggest that wrapping with SDS has another effect in addition to the dielectric screening effect. The behavior of the additional shifts is similar to the shift induced by uniaxial stress, reported by Arnold et al.[18] At present, it still remains an issue whether such uniaxial strain is induced in the SWNTs by wrapping with surfactants or not. A charging effect due to SDS, which is an anionic surfactant, should also be considered. Further study is necessary to understand the effects of surfactants on optical transition energies in SWNTs.

Finally, we note the broadening of the PL spectra in liquids. The representative spectra are shown in Fig. 5. The peaks show a broadening, in addition to the redshift with increasing $\epsilon_{\text{env}}$. The linewidth increased from 23 meV in air to 40 meV in acetonitrile for (9,7) SWNTs. This linewidth broadening is probably attributed to inhomogeneous broadening due to local fluctuations of $\epsilon_{\text{env}}$ on the nano-scale dimension. The dielectric constants we used in Table I are macroscopic values. On such small dimensions as exciton diameters of a few nm[24], the size of molecules of the organic solvents would be considerable so that the local $\epsilon_{\text{env}}$ would fluctuate depending on the number and orientation of the organic molecules. This would result in inhomogeneous broadening of the PL spectrum.

## IV. SUMMARY

In summary, dielectric screening effects due to the environment around SWNTs on exciton transition energies in the SWNTs have been studied quantitatively by means of PL and the excitation spectroscopies. We varied the $\epsilon_{\text{env}}$ from 1.0 to 37 by immersing the samples with SWNTs bridging over trenches in various organic solvents with different $\epsilon_{\text{env}}$s. With increasing $\epsilon_{\text{env}}$, both $E_{11}$ and $E_{22}$ showed a redshift by 33~49 meV for $E_{11}$ and 26~30 meV for $E_{22}$, and a tendency to saturate at $\epsilon_{\text{env}} \sim 5$, without a significant $(n,m)$ dependence. The redshift can be explained by dielectric screening of repulsive electron-electron energy. The $\epsilon_{\text{env}}$ dependence of $E_{11}$ and $E_{22}$ were expressed by a simple empirical equation with a power law in $\epsilon_{\text{env}}$. The equivalent $\epsilon_{\text{env}}$ of SDS-wrapped SWNTs was estimated to be $\sim 2$. It was suggested that the effect of wrapping SWNTs with SDS was not only a dielectric screening effect, but also another effect which caused an energy shift like a uniaxial-stress-induced shift.

---

TABLE I: $\epsilon_{\text{env}}$ of air and various liquids used in this study.

| solvent | $\epsilon_{\text{env}}$ |
|---|---|
| air | 1.0 |
| hexane | 1.9 |
| chloroform | 4.8 |
| ethyl acetate | 6.0 |
| dichloromethane | 9.0 |
| acetone | 21 |
| acetonitrile | 37 |

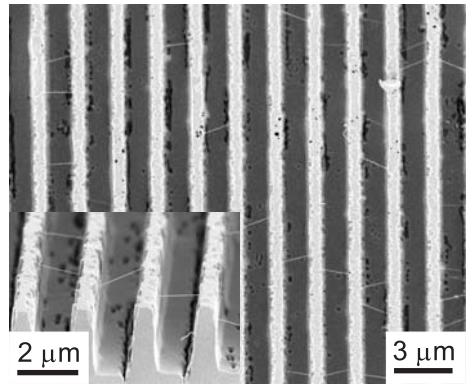

FIG. 1: Plan-view and bird-view (inset) SEM images of SWNTs bridging over trenches on a quartz substrate. Pt thin film was deposited on the sample in order to avoid charge up of the insulating substrate and to observe individual SWNTs. By optimizing growth condition, individual SWNTs can be obtained.

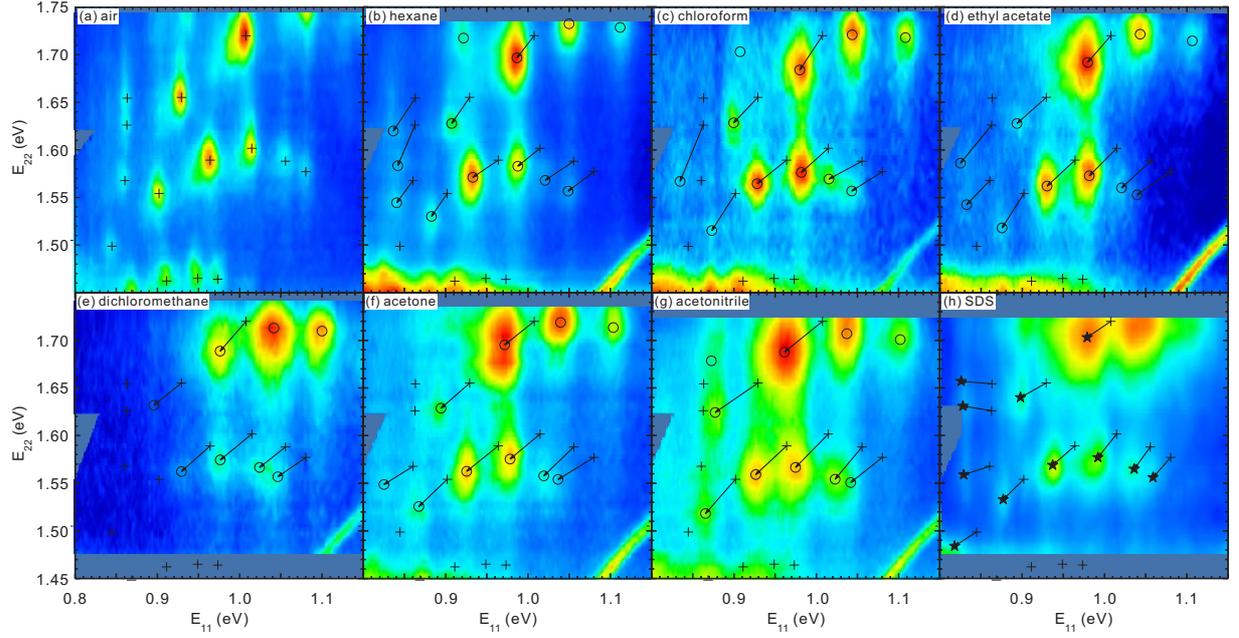

FIG. 2: PL contour maps for SWNTs; (a) in air ($\epsilon_{\text{env}} = 1.0$), (b) hexane ($\epsilon_{\text{env}} = 1.9$), (c) chloroform ($\epsilon_{\text{env}} = 4.8$), (d) ethyl acetate ($\epsilon_{\text{env}} = 6.0$), (e) dichloromethane ($\epsilon_{\text{env}} = 9.0$), (f) acetone ($\epsilon_{\text{env}} = 20.7$), and (g) acetonitrile ($\epsilon_{\text{env}} = 37.5$), and (h) SDS/D$_2$O solution. The crosses and open circles represent peak positions in air and in the liquid. In (h), stars represent the peak positions of SDS-wrapped SWNTs reported by Weisman *et al*[3]. The PL intensity color schemes are linear, but different scaling factors were used for each maps. The emission and excitation maxima correspond to $E_{11}$ and $E_{22}$, respectively.

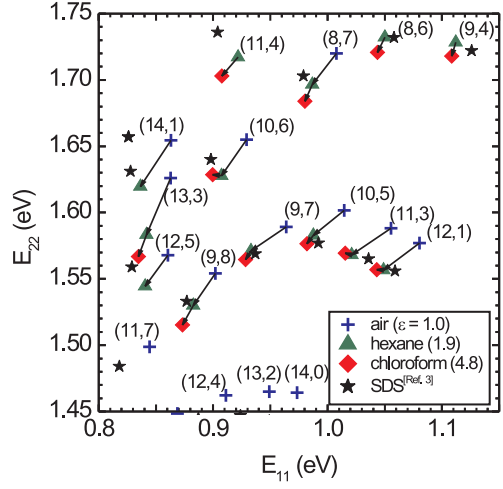

FIG. 3: $E_{22}$ versus $E_{11}$ plots in air ($\epsilon_{\text{env}} = 1.0$), hexane ($\epsilon_{\text{env}} = 1.9$), chloroform ($\epsilon_{\text{env}} = 4.8$). The stars represent the peak positions of SDS-wrapped SWNTs[3]. Both $E_{11}$ and $E_{22}$ show redshifts with increasing $\epsilon_{\text{env}}$ with a small $(n,m)$ dependence. The peak positions of SDS-wrapped SWNTs deviate from the line of dielectric screening effect.

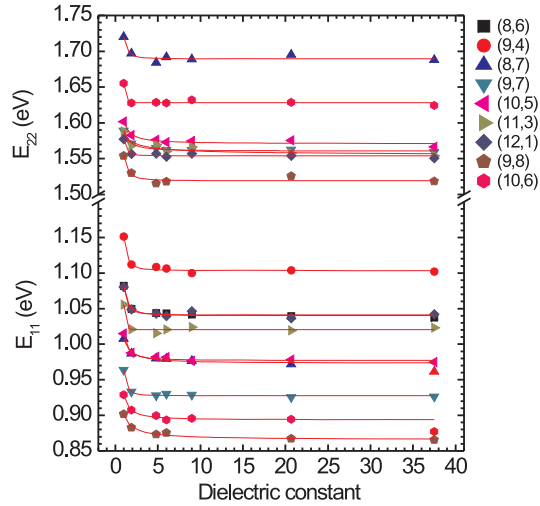

FIG. 4: $\epsilon_{\text{env}}$ dependences of $E_{11}$ and $E_{22}$ of various $(n,m)$ SWNTs. The solid lines are fitting curves given by an empirical expression with power law in $\epsilon_{env}$.

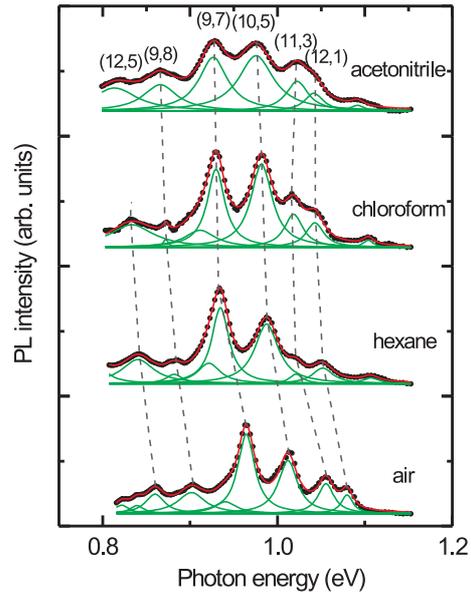

FIG. 5: PL spectra in various liquids. With increasing $\epsilon_{env}$, the emission spectra show linewidth broadening in addition to redshift.